\documentclass[conference]{IEEEtran}
\IEEEoverridecommandlockouts

\usepackage{cite}
\usepackage{amsmath,amssymb,amsfonts}
\usepackage{algorithmic}
\usepackage{graphicx}
\usepackage{textcomp}
\usepackage{xcolor}
\def\BibTeX{{\rm B\kern-.05em{\sc i\kern-.025em b}\kern-.08em
    T\kern-.1667em\lower.7ex\hbox{E}\kern-.125emX}}
\begin{document}

\title{Human-AI Collaborative Game Testing with Vision Language Models\\[0.2em]
{\normalfont\large\itshape An Experiment Report}
\thanks{\IEEEauthorrefmark{1}Corresponding author.}
}

\author{\IEEEauthorblockN{Boran Zhang}
\IEEEauthorblockA{
\textit{Zhengzhou University}\\
Zhengzhou, China \\
zhangboran12138@gs.zzu.edu.cn}
\and
\IEEEauthorblockN{Muhan Xu}
\IEEEauthorblockA{
\textit{University of the Arts London}\\
London, United Kingdom \\
m.xu0920221@arts.ac.uk}
\and
\IEEEauthorblockN{Zhijun Pan\IEEEauthorrefmark{1}}
\IEEEauthorblockA{
\textit{Royal College of Art}\\
London, United Kingdom \\
zhijun.pan@network.rca.ac.uk}

}

\maketitle

\begin{abstract}
As modern video games become increasingly complex, traditional manual testing methods are proving costly and inefficient, limiting the ability to ensure high-quality game experiences. While advancements in Artificial Intelligence (AI) offer the potential to assist human testers, the effectiveness of AI in truly enhancing real-world human performance remains underexplored. This study investigates how AI can improve game testing by developing and experimenting with an AI-assisted workflow that leverages state-of-the-art machine learning models for defect detection. Through an experiment involving 800 test cases and 276 participants of varying backgrounds, we evaluate the effectiveness of AI assistance under four conditions: with or without AI support, and with or without detailed knowledge of defects and design documentation. The results indicate that AI assistance significantly improves defect identification performance, particularly when paired with detailed knowledge. However, challenges arise when AI errors occur, negatively impacting human decision-making. Our findings show the importance of optimizing human-AI collaboration and implementing strategies to mitigate the effects of AI inaccuracies. By this research, we demonstrate AI's potential and problems in enhancing efficiency and accuracy in game testing workflows and offers practical insights for integrating AI into the testing process.
\end{abstract}

\begin{IEEEkeywords}
Artificial Intelligence, Game Testing, Human-AI Collaboration, Test Automation, Error Detection, Process Optimization.
\end{IEEEkeywords}

\section{Introduction}

The rapid growth of the gaming industry has led to the development of increasingly complex and immersive video games, characterized by dynamic mechanics, interactive narratives, and expansive open worlds \cite{levy2009game,barr2007video,liang2023ag3,chueca2024consolidation}. Ensuring the functionality and quality of such intricate games requires comprehensive testing. However, traditional manual game testing approaches are becoming insufficient due to the scale and complexity of modern games. These methods are time-consuming, resource-intensive, and prone to human error, reducing their effectiveness in identifying a broad range of defects \cite{politowski2021survey,pfau2017automated,rahman2024quantizing}.

While average gamers can serve as video game playtesters, Artificial Intelligence (AI) has emerged as a promising solution to enhance game testing workflows by automating defect detection processes and reducing reliance on manual efforts. Early AI approaches in game testing focused on rule-based automation to manage repetitive tasks like regression testing. However, these systems struggled to adapt to the emergent gameplay dynamics and player behaviors found in complex gaming environments \cite{dhavachelvan2006new,levy2009game,ramadan2014development}. Recent advancements in machine learning, particularly reinforcement learning, have enabled AI systems to autonomously simulate player interactions and learn from game environments, improving both the realism and robustness of test cases \cite{taesiri2020video,durelli2019machine,paduraru2021automated, pfau2017automated,azizi2024astrobug}. These AI systems are capable of detecting a wide spectrum of defects, ranging from visual glitches to gameplay inconsistencies, enhancing the depth and efficiency of game testing.

A key advancement in AI-driven testing is the integration of multimodal inputs—visual, audio, and textual data—which allows AI systems to comprehensively analyze gameplay environments. By leveraging these diverse data types, AI systems can detect graphical anomalies, such as texture misalignments and rendering errors, while simulating player behaviors to uncover gameplay issues \cite{paduraru2024automated,liang2023ag3,rahman2024quantizing,liu2022inspector,senchenko2022supernova,arcuri2017restful, ariyurek2019automated}. However, while AI systems offer enhanced capabilities, they also present challenges, particularly in misinterpreting complex game elements, which can lead to inaccurate defect identification or even hallucinations—where the AI erroneously flags non-defective elements as bugs. These hallucinations, compounded by human testers' reliance on AI, can further complicate the testing process \cite{dwivedi2021artificial,cabitza2024explanations,xu2021human,rawte2023survey,zhang2023siren}. Solutions to mitigate AI hallucinations are actively being explored to enhance the reliability of AI systems \cite{sun2023contrastive,chen2024honest}.

Current AI-driven game testing approaches can be classified into two main categories: rule-based automation and AI-driven perception systems \cite{stahlke2018usertesting}. Rule-based systems, which simulate player actions by interacting directly with the game's source code, require substantial technical expertise and may interfere with game performance during testing \cite{liang2023ag3,jintawatsakoon2023visual}. In contrast, AI-driven perception systems utilize machine vision and multimodal models to detect visual defects, such as rendering errors, and analyze gameplay mechanics \cite{liu2022inspector,paduraru2021automated}. While effective, these systems may still struggle with unpredictable player behavior and nuanced gameplay scenarios, areas where human testers excel \cite{azizi2024astrobug,senchenko2022supernova,sestini2022automated}.

Despite the advances in AI, the extent to which AI enhances human performance in real-world game testing, particularly under various conditions, remains underexplored. This study aims to address this gap by investigating the impact of AI on human testers' performance across different levels of knowledge and experience, and by exploring how AI-generated insights can improve defect detection while managing risks like hallucinations.

The contributions of our research are as follows:
\begin{itemize}
    \item \textbf{AI-Assisted Game Testing Workflow}: We present an AI-assisted workflow that integrates machine learning to detect defects in game visuals, user interfaces, and gameplay mechanics.
    \item \textbf{Empirical Insights into Human-AI Collaboration}: Through a comprehensive user study, we provide empirical evidence demonstrating how AI enhances game testers under various conditions, especially when paired with detailed knowledge and human oversight.
    \item \textbf{Analysis of AI Hallucinations and Error Impact}: We analyze how AI hallucinations and other errors impact human testers' performance, offering strategies to mitigate over-reliance and reduce the effects of AI inaccuracies.
    \item \textbf{Practical Guidelines for AI Integration}: Based on the findings, we propose practical recommendations for incorporating AI into game testing workflows, emphasizing the balance between AI assistance and human judgment.
\end{itemize}

\section{Related Work}

\subsection{Multimodal AI}

Multimodal AI involves integrating diverse data types—such as visual, audio, and textual inputs—into AI systems, enabling more robust and comprehensive analysis \cite{liu2022inspector,senchenko2022supernova}. In the context of game testing, multimodal AI facilitates the detection of a wide array of defects, from visual glitches to inconsistencies in gameplay and narrative \cite{paduraru2024automated,liang2023ag3,meszaros2007xunit}. This approach allows AI systems to assess multiple dimensions of a game simultaneously, improving the breadth and depth of testing.

Recent advancements in deep learning models designed for multimodal inputs have significantly enhanced AI's real-time, multidimensional capabilities in game testing \cite{openai2024gpt4o}. These models can process complex game environments, evaluating everything from graphical fidelity to narrative coherence \cite{pfau2017automated,paduraru2021automated,he2016deep,krizhevsky2017imagenet}. However, integrating multiple data types introduces challenges, including the need for large and diverse datasets and the computational complexity required for real-time analysis \cite{rahman2024quantizing,arcuri2017restful}. Moreover, the "black-box" nature of many deep learning models raises concerns about explainability, which can hinder human testers' trust in AI systems \cite{prasetya2022agent,cabitza2024explanations}.

Another notable challenge with multimodal AI systems is hallucinations—where the AI misinterprets game elements and flags non-existent bugs \cite{rawte2023survey}. Addressing this issue is crucial for improving AI reliability. Recent research has explored techniques like contrastive learning and model fine-tuning to reduce hallucinations and improve AI accuracy in defect detection \cite{sun2023contrastive,chen2024honest,zhang2023siren}. These efforts are essential to ensuring that AI-assisted systems remain trustworthy and efficient in complex game testing environments.

\subsection{Human-AI Collaboration}

Human-AI collaboration in software testing combines the strengths of automated AI systems and human testers' intuitive judgment. AI excels at automating repetitive tasks, such as generating test cases and identifying specific types of defects, thereby enhancing the efficiency and scalability of testing processes \cite{durelli2019machine,chen2021glib}. Human testers, on the other hand, contribute critical thinking, intuition, and the ability to manage complex or edge-case scenarios that are often difficult to automate \cite{norman2013design,shneiderman2016designing,dix2004human}.

In game testing, which often involves dynamic environments and unpredictable player behaviors, balancing human-AI collaboration is essential for thorough defect detection \cite{levy2009game,hartson2018ux}. Studies have shown that over-reliance on AI systems can lead testers to overlook subtle defects or unquestioningly accept incorrect AI outputs, underscoring the importance of workflows that allow for human oversight and critical assessment \cite{xu2021human,cabitza2024explanations}. To mitigate these issues, explainable AI (XAI) techniques have been explored to improve the transparency of AI systems, especially when errors, such as hallucinations, occur \cite{lazar2017research,dwivedi2021artificial,vodrahalli2022uncalibrated}.

Recent studies indicate that machine input can heavily influence human decision-making, especially under cognitive limitations \cite{boyaci2024human}. Trust between human testers and AI systems is vital for effective collaboration \cite{gebru2022review}, and dissonance between human and machine understanding can cause inefficiencies \cite{zhang2019dissonance}. Well-designed interfaces should enable testers to adjust or override AI decisions, minimizing the impact of AI errors \cite{harms2019automated,varvaressos2017automated}. Improved AI training methods and handling uncertainty are essential to mitigate these issues \cite{chen2024honest}.

\subsection{AI-Assisted Game Testing}

AI-assisted game testing has evolved from rule-based automation to more advanced machine learning and reinforcement learning techniques \cite{dhavachelvan2006new,levy2009game,rani2023deep,albaghajati2020video, ariyurek2019automated}. Deep learning architectures, particularly convolutional neural networks (CNNs), have proven effective in detecting visual defects, such as texture misalignments and rendering errors \cite{liu2022inspector}. Systems like AstroBug \cite{azizi2024astrobug} and SUPERNOVA \cite{senchenko2022supernova} utilize these models to analyze gameplay footage in real time, offering scalable solutions that outperform manual testing in several aspects.

Reinforcement learning, which simulates diverse player behaviors, has also been successfully applied to uncover bugs related to game physics, balance, and interaction dynamics, often missed by traditional testing methods \cite{paduraru2021automated,prasetya2022agent,jintawatsakoon2023visual,nantes2008framework}. This approach allows AI to explore edge cases and emergent gameplay behaviors, improving the thoroughness of game testing \cite{pfau2017automated,paduraru2024automated}.

Nevertheless, AI-assisted game testing faces limitations. AI systems may struggle with the unpredictability of complex game environments, potentially missing nuanced defects that require human interpretation \cite{varvaressos2017automated,dwivedi2021artificial}. Human oversight is therefore critical for validating AI-generated reports and refining models to better handle complex scenarios \cite{paduraru2022rivergame,liu2022inspector}.

\section{Methodology}

\subsection{Game Testing Workflow}\label{GTP}

The primary objective of our research is to investigate how AI can effectively assist human testers in identifying defects in video games, focusing on integrating AI-driven automation with human oversight. We employed two principle workflows: a traditional manual testing process and an AI-assisted testing process.

\begin{figure}
    \centering
    \includegraphics[width=0.98\linewidth]{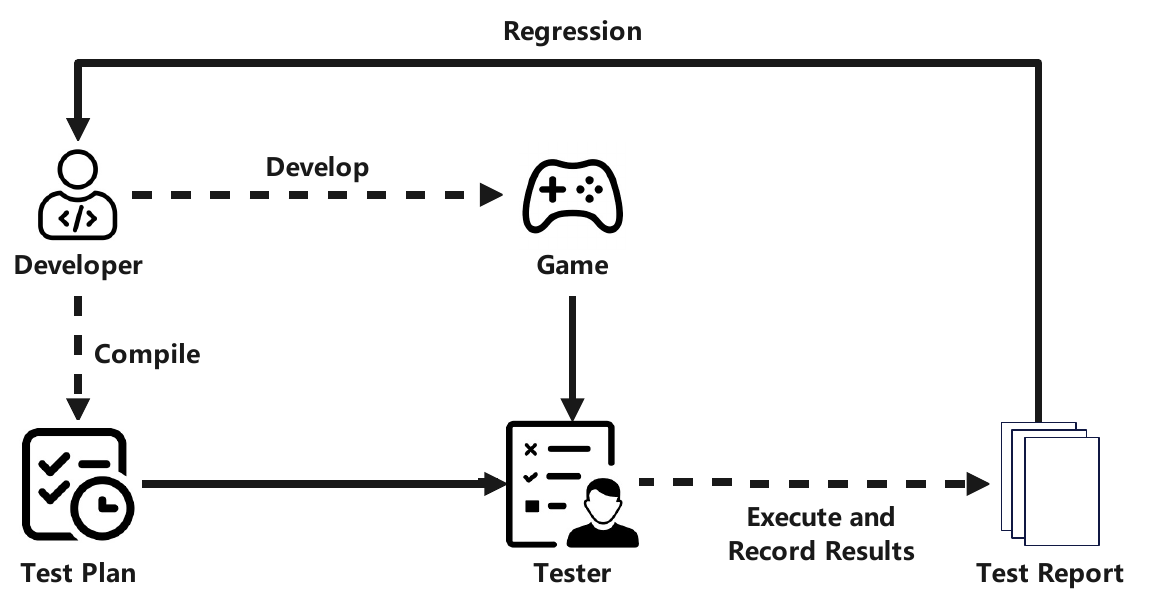}
    \caption{Traditional Game Testing Workflow (Simplified)}
    \label{fig1}
\end{figure}

In the traditional game testing workflow (illustrated in Fig \ref{fig1}), the process starts with game development and the creation of a detailed test plan. This test plan outlines the objectives, requirements, and scope of testing, covering both game mechanics and graphical elements \cite{ballou2024basic}. Human testers manually execute the test cases, aiming to identify defects such as graphical glitches, gameplay inconsistencies, and physics issues. Defects identified by testers are documented in a test report, which is then forwarded to developers for resolution. After the resolution, regression testing is conducted to ensure no new issues have been introduced. This workflow, although thorough, is resource-intensive, time-consuming, and prone to human error, particularly for large and complex games \cite{politowski2021survey,schultz2016game}.

\begin{figure}
    \centering
    \includegraphics[width=0.98\linewidth]{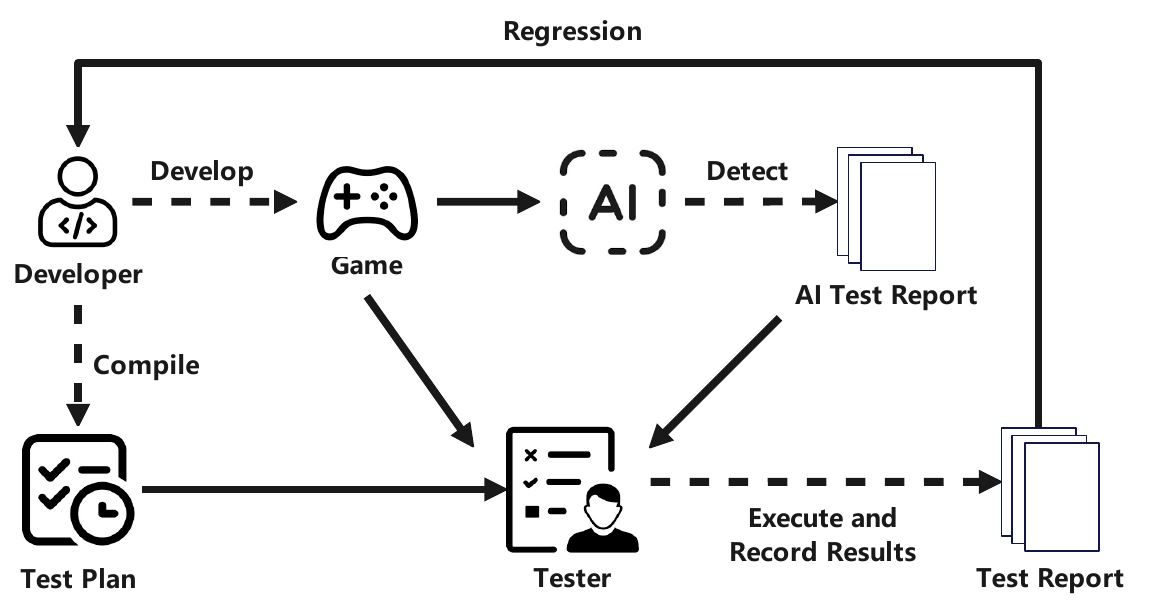}
    \caption{AI-Assisted Game Testing Workflow (Simplified)}
    \label{fig2}
\end{figure}

In the AI-assisted testing workflow (illustrated in Fig \ref{fig2}), the process begins similarly with game development and test planning. However, during the test execution phase, an AI component captures in-game screenshots and analyzes them using state-of-the-art models. The AI system detects visual defects such as texture misalignments, clipping errors, and UI inconsistencies, and compiles these findings into an initial test report for human reviewers. Human testers then validate the AI-identified defects and focus on more complex issues like gameplay mechanics and physics behaviors, which may not be easily detectable by AI. This hybrid approach enables AI to handle repetitive, surface-level defect detection, allowing human testers to spend more time on in-depth analysis. The final test report combines both AI-generated and human-verified results, which are submitted to developers for resolution. 

This hybrid workflow represents our core research framework: an integration of automated detection with human validation in game testing. The setup allows us to explore the effectiveness of such a system in improving testing efficiency, accuracy, and overall human performance.

\subsection{Experiment Setup}

Our experiment was designed to validate the AI-assisted game testing system's reliability and examine the effects of human-AI collaboration in identifying game defects. The experiment setup consists of the following components: the system, test cases, execution procedures, and scoring mechanisms.

\subsubsection{System}
The AI-assisted system is based on the state-of-the-art GPT-4o model \cite{openai2024gpt4o}, which was employed to process a diverse dataset of video game visuals, including different defect types and gameplay elements. The AI component functions by processing game screenshots, generating descriptions of the game environment, and detecting a set of predefined defect categories (see Table \ref{tab2}). Human participants interact with this system during the testing phases, working the AI's outputs to identify game defects. The system's performance metrics include the accuracy of visual descriptions and defect identifications, with an 87.5\% accuracy rate for scene descriptions and a 97.1\% accuracy rate for defect detection under controlled conditions.

\subsubsection{Test Cases and Execution}
The experiment involved a total of 800 test cases, divided equally across four experimental conditions: No Knowledge, No AI Assistance (Manual); Knowledge, No AI Assistance (KA); No Knowledge, With AI Assistance (AIA); and Knowledge, With AI Assistance (KA+AIA). Each condition consisted of 200 test cases, and each participant took a total of 40 random cases—10 from each condition. This ensured that all participants experienced every possible combination of AI assistance and the availability of knowledge.

In this experiment, ``knowledge" refers to detailed information generated by the AI system, including explanations of potential defects identified by the AI (when AI involves), as well as the game design documentation related to the specific test cases. This knowledge helps participants understand the types of defects that may occur and provides context from the game’s design to assist in defect identification.

Each test case involved the identification of defects from game images, with participants required to identify up to two defects per case. Defects were selected from four primary categories: graphical rendering errors, user interface (UI) errors, scene and object consistency errors, and object and physics behavior errors, which were further subdivided into specific types as described in Table \ref{tab2}. While most test cases contains at least one defect to be identifies, around 10\% of all cases are bug-free.

Test execution followed a structured sequence, ensuring participants experienced both AI-assisted and non-AI-assisted conditions, as well as conditions with and without the detailed knowledge provided by the AI system.

\subsubsection{Scoring Mechanism}
Participants were scored based on their ability to correctly identify defects within each test case. The scoring system was designed to allocate a maximum of 10 points per test case, with 5 points awarded for each correctly identified defect. Incorrect or missed defect identifications resulted in no points for that particular defect. Bug-free cases come with 10 points initially. If the participant identified bugs for bug-free cases, there will be no point. Therefore, each participant could score a maximum of 400 points, with the average score recorded as 211 points and the highest as 395. This scoring mechanism enabled a comparative analysis of participants' performance across different test conditions.

\renewcommand{\arraystretch}{1.2} 

\begin{table}
   \caption{Demographics of Human Participants}
    \centering
    \begin{tabular}{|c|c|} \hline 
         & M(SD) or n(\%)\\ \hline 
         \multicolumn{2}{|l|}{\textbf{Age}}\\ \hline 
         Under 18& 1(0.4\%)\\ \hline 
         18–24 years& 153 (55.4\%)\\ \hline 
         25–30 years& 104 (37.7\%)\\ \hline 
         Over 30 years& 18 (6.5\%)\\ \hline 
         \multicolumn{2}{|l|}{\textbf{Gaming Experience}}\\ \hline 
         Yes&269 (97.5\%)\\ \hline 
         No&7 (2.5\%)\\ \hline 
         \multicolumn{2}{|l|}{\textbf{Education Level}}\\ \hline 
         High School or below&17 (6.2\%)\\ \hline 
         Associate Degree&84(30.4\%)\\ \hline 
         Bachelor’s Degree&160 (58\%)\\ \hline 
         Master’s Degree or above&15 (5.4\%)\\ \hline
    \end{tabular}
    \label{tab1}
\end{table}

\subsection{Participants}\label{P}

\paragraph{Human Participants}: A total of 276 human participants were recruited via an online platform, with the average test duration being 6 minutes and 5 seconds. Among the participants, 55.4\% were between 18 and 24 years old, and 97.5\% reported prior gaming experience. For a detailed breakdown of participant demographics, see Table \ref{tab1}.
\paragraph{AI Component}: The AI component functions as an additional participant in the testing process, utilizing advanced machine vision techniques to analyze and detect defects. Its performance, as mentioned, was rigorously evaluated for accuracy in both scene description and defect identification.

\subsection{Tasks}\label{T}

The primary task for participants was to identify defects in the provided game visuals and classify them based on the defined categories. Defects were categorized into graphical rendering errors, UI errors, scene and object consistency errors, and object and physics behavior errors\cite{coppola2024know}, as detailed in Table \ref{tab2}. Participants were required to complete the task under four different experimental conditions.

\subsection{Conditions}\label{C}

Our experiment was structured into four testing conditions that combined the AI system with the availability of detailed defect information and design documentation (knowledge):

\begin{itemize}
\item \textbf{No Knowledge, No AI Assistance (Manual)}: Participants relied solely on observation without any detailed defect information or AI assistance.
\item \textbf{Knowledge, No AI Assistance (KA)}: Participants were provided with detailed explanations of potential defects identified by the AI, along with game design documentation related to the test cases. However, they performed the tests without direct AI assistance during execution.
\item \textbf{No Knowledge, With AI Assistance (AIA)}: Participants used the AI system for defect identification but did not have access to detailed defect information or game design documentation.
\item \textbf{Knowledge, With AI Assistance (KA+AIA)}: Participants had both the AI system’s assistance and access to detailed explanations of potential defects, as well as game design documentation for the test cases.
\end{itemize}

\subsection{Procedure}

The experimental procedure followed these steps:

\begin{itemize}
\item \textbf{Pre-Test Briefing}: Participants were briefed on the experiment’s content and procedure.
\item \textbf{Test Execution}: Participants evaluated 10 randomly assigned game screenshots in each of the four conditions (Manual, KA, AIA, KA+AIA), following a predefined sequence. 
\item \textbf{Data Collection}: Participants’ responses, including the difficulty of the game screenshots and the AI system's evaluations, were recorded.
\item \textbf{Post-Test Feedback}: Participants provided feedback on their experience, including their understanding of the AI system and defect identification process.
\item \textbf{Incentive Mechanism}: Participants were incentivized with monetary rewards based on their performance, with bonuses for higher accuracy scores.
\end{itemize}

\renewcommand{\arraystretch}{1.5} 

\begin{table*}[t]
\caption{Classification and Definitions of Game Bugs}
\centering
\begin{tabular}{|p{2.5cm}|l|p{7cm}|p{2cm}|}
\hline
\textbf{Category} & \textbf{Bug Type} & \textbf{Definition} & \textbf{References} \\
\hline
{Graphical Rendering Errors} 
& Missing Textures & Failure of texture assets to load properly, leading to objects being rendered with placeholder colors or default textures. & {\cite{sestini2022automated}, \cite{varvaressos2017automated}, \cite{prasetya2022agent}, \cite{paduraru2022rivergame}, \cite{chen2021glib}, \cite{jiang2021droidgamer}} \\

\cline{2-3} 
& Lighting/Shadow Artifacts & Anomalies in lighting or shadow calculations, such as inaccurate shadow placement or improper light distribution, affecting scene realism. & \multicolumn{1}{l|}{} \\

\cline{2-3}
& Clipping Issues & Objects that intersect or pass through other objects or environmental boundaries, violating expected spatial constraints (e.g., objects passing through walls). & \multicolumn{1}{l|}{} \\

\cline{2-3}
& Resolution or Scaling Anomalies & Visual distortion due to incorrect game resolution or object scaling, leading to blurry textures or disproportionate object sizes. & \multicolumn{1}{l|}{} \\

\cline{2-3}
& Texture Tiling Artifacts & Visual seams or repetition artifacts when the same texture is tiled across large surfaces, disrupting visual continuity. & \multicolumn{1}{l|}{} \\

\hline
{User Interface (UI) Errors} 
& UI Element Overlap & Overlapping or improperly aligned UI elements (e.g., buttons, text), resulting in reduced clarity and hindered user interaction. & {\cite{chen2021glib}, \cite{paduraru2022rivergame}} \\

\cline{2-3}
& Display Artifacts & Rendering issues within the UI, such as misaligned elements or text, negatively affecting interface usability and information accessibility. & \multicolumn{1}{l|}{} \\

\hline
{Scene and Object Consistency Errors} 
& Missing or Deformed Objects & Incomplete or improperly rendered in-game objects, leading to visual deformities or missing elements within the game environment. & {\cite{pfau2017automated}, \cite{rani2023deep}, \cite{taesiri2020video}} \\

\cline{2-3}
& Scene Loading Errors & Incomplete or misaligned loading of the game environment, such as unrendered terrain or structures, resulting in visual dissonance. & \multicolumn{1}{l|}{} \\

\hline
{Object and Physics Behavior Errors} 
& Abnormal Object Movement & Unnatural object or character movement, such as erratic position shifts or paths that contradict expected physical behavior. & {\cite{paduraru2022rivergame}, \cite{liu2022inspector}} \\

\cline{2-3}
& Physics Engine Errors & Errors in the physics simulation, where objects behave contrary to physical laws, such as passing through solid objects or bouncing incorrectly. & \multicolumn{1}{l|}{} \\

\hline No Bug & N/A & The game environment functions as expected with no detectable graphical, UI, consistency, or physics-related issues. & N/A \\

\hline

\end{tabular}
\label{tab2}
\end{table*}

\subsection{Data Analysis Method}\label{DA}
\subsubsection{Data Preprocessing}\label{DP}
To ensure the validity and accuracy of the data, the following preprocessing steps were conducted:
\begin{itemize}
\item \textbf{Exclusion of Anomalous Times:} Each participant's response duration was recorded, with an average analysis time per case exceeding 8 seconds. Therefore, any response time under 300 seconds was considered indicative of insufficient engagement. A minimum threshold of 300 seconds was established, and all records falling below this threshold were excluded.
\item \textbf{Handling of Score Anomalies:} Total scores for each participant underwent Z-score normalization. Records with absolute Z-scores greater than 3 were classified as outliers and removed.
\end{itemize}




\subsubsection{Calculation of Accuracy}

To evaluate participants' performance under the four testing conditions—Manual (M), Knowledge Assistance (KA), AI Assistance (AIA), and Knowledge plus AI Assistance (KA+AIA)—we calculated the accuracy of defect identification.

Let \( K = \{ \text{M}, \text{KA}, \text{AIA}, \text{KA{+}AIA} \} \) denote the set of testing conditions, \( C \) the set of all test cases, and \( P \) the set of all participants. Each participant \( i \in P \) completed a subset of test cases \( C_{i,k} \subseteq C \) under condition \( k \in K \). For each test case \( j \) under condition \( k \), let \( P_{j,k} \subseteq P \) be the set of participants who took test case \( j \). The score achieved by participant \( i \) on test case \( j \) under condition \( k \) is denoted by \( S_{i,j,k} \), and \( S_{\text{max}} \) is the maximum possible score per test case.

\paragraph{Accuracy per Participant and Condition}

The accuracy \( A_{i,k} \) for participant \( i \) under condition \( k \) is calculated as:

\begin{equation}
A_{i,k} = \frac{\sum_{j \in C_{i,k}} S_{i,j,k}}{|C_{i,k}| \times S_{\text{max}}}
\end{equation}

\paragraph{Average Accuracy per Condition}

The average accuracy \( \bar{A}_k \) for condition \( k \) is:

\begin{equation}
\bar{A}_k = \frac{\sum_{i \in P_k} \sum_{j \in C_{i,k}} S_{i,j,k}}{\sum_{i \in P_k} |C_{i,k}| \times S_{\text{max}}}
\end{equation}

where \( P_k = \{ i \in P \mid C_{i,k} \neq \emptyset \} \) is the set of participants who completed any test cases under condition \( k \).

\paragraph{Accuracy per Test Case and Condition}

The accuracy \( A_{j,k} \) for test case \( j \) under condition \( k \) is:

\begin{equation}
A_{j,k} = \frac{\sum_{i \in P_{j,k}} S_{i,j,k}}{|P_{j,k}| \times S_{\text{max}}}
\end{equation}

\paragraph{Accuracy per Scenario and Condition}

For a scenario \( s \) (a set of test cases within a specific category), the accuracy \( A_{s,k} \) under condition \( k \) is:

\begin{equation}
A_{s,k} = \frac{\sum_{j \in C_s} \sum_{i \in P_{j,k}} S_{i,j,k}}{\sum_{j \in C_s} |P_{j,k}| \times S_{\text{max}}}
\end{equation}

where \( C_s \subseteq C \) is the set of test cases in scenario \( s \).

\paragraph{Explanation}

This methodology accounts for varying numbers of participants per test case and condition. By aggregating scores and normalizing them with the appropriate maximum possible scores, we ensure accurate calculation of accuracy rates despite differences in participant assignments and test case distributions across conditions and scenarios. This allows for a comprehensive evaluation of performance across different testing conditions and scenarios.

\subsubsection{Challenges Faced by Testers}\label{CFT}
Through participant interviews, we identified several issues that may have hindered accurate judgment during the testing process, and we have classified the challenging test cases as follows:
\begin{itemize}
\item \textbf{Low Visibility (LV):} In low-visibility conditions, participants often struggle to identify potential graphical output errors, making defect detection challenging. This situation may lead to incorrect classifications as Lighting/Shadow Artifacts.
\item \textbf{Multiple Defects (MD):} Many game issues arise from the interplay of multiple defects, and participants may prioritize more apparent defects while overlooking others, leading to incomplete defect assessments.
\item \textbf{Subtle Defects (SD):} Subtle defects, such as missing structures indicated only by minimal hints or minor object misbehavior, are often overlooked or mistaken for normal occurrences, resulting in erroneous judgments.
\item \textbf{Ambiguous Boundaries (AB): } Different defect types may exhibit similar characteristics. For instance, scene loading errors in object consistency can sometimes resemble texture loss in graphical rendering errors, leading to misjudgments by participants.
\end{itemize}

\subsubsection{Potential AI Errors}
The output from the AI-assisted system is categorized into scene descriptions and defect assessments. During the testing process, we observed potential errors in the AI system that could influence participants' results and classified test cases accordingly. The main types of errors include:
\begin{itemize}
\item \textbf{Descriptive Errors (DE):} When the AI system inaccurately describes a game screenshot, participants may derive incorrect conclusions during defect analysis.
\item \textbf{Judgment Errors (JE):} Instances of misclassification in defect type recognition by the AI system may lead participants to rely excessively on the AI's conclusions, resulting in erroneous judgments.
\item \textbf{Correct Description, Judgment Error (CDJE):} If the AI system accurately describes the scene but makes incorrect judgments, participants may trust the AI's assessment and consequently make errors. This scenario frequently occurs when defect boundaries are unclear or when defects are non-unique.
\item \textbf{Both Description and Judgment Errors (DEJE):} When the AI system fails in both description and judgment, participants might correct erroneous judgments upon noticing discrepancies between the AI's description and the visuals, or they may blindly trust the AI’s seemingly logical conclusions, leading to mistakes.
\end{itemize}

\subsubsection{AI Hallucination Cases}
AI hallucinations—when the system incorrectly detects defects—presented additional challenges for testers:
\begin{itemize}
    \item \textbf{False Negative Detections (FND):} In cases where defects were subtle or nuanced, the AI sometimes failed to detect any issues, incorrectly assessing the game visuals as defect-free. This mirrored the difficulty human testers experience in traditional testing, potentially leading to incomplete or inaccurate test reports.
    
    \item \textbf{False Positive Detections (FPD):} When game visuals were highly stylized or contained elements that defied conventional design patterns (e.g., special effects or exaggerated color schemes), the AI frequently misidentified these as defects. Such misjudgments often occurred due to insufficient contextual information provided about the game's design, causing the AI to flag intended artistic choices as errors.
\end{itemize}

\section{Experiment Results and Analysis}

\subsection{Summary of Key Findings}

The results of this study indicate that AI-assisted systems significantly enhance the effectiveness of human testers in identifying game defects, especially under challenging conditions. The integration of AI assistance with detailed knowledge of defects—including information generated by AI and access to game design documentation—yielded the highest effectiveness in bug detection. However, the study also uncovered unexpected patterns regarding the impact of AI errors and hallucinations on human decision-making, highlighting the complexities of human-AI collaboration in game testing.

\begin{figure}
    \centering
    \includegraphics[width=1\linewidth]{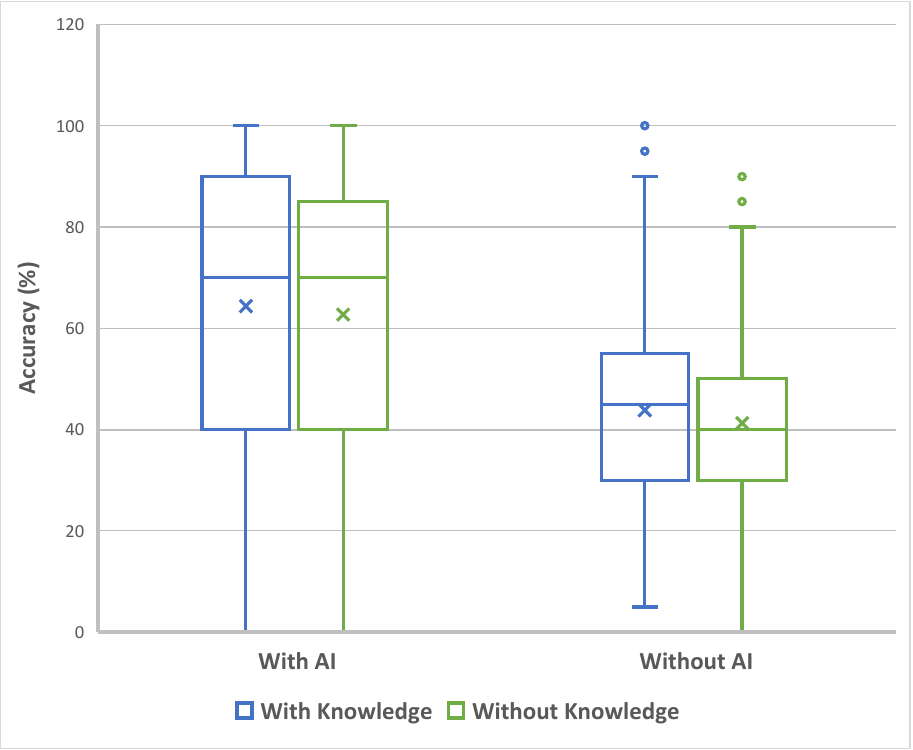}
    \caption{Test accuracy distribution. This image illustrates the maximum, average, upper quartile, median, lower quartile, and minimum accuracies of participants under different testing conditions.}
    \label{fig3}
\end{figure}

\subsection{Comparison of Effectiveness Across Conditions}\label{CSWD}

To evaluate the impact of AI assistance and knowledge provision on defect detection accuracy, we compared participants' performance across the four experimental conditions: Manual (M), Knowledge Assistance (KA), AI Assistance (AIA), and Knowledge plus AI Assistance (KA+AIA). As shown in Figure~\ref{fig3}, AI assistance significantly improved defect detection accuracy. Specifically, participants in the AIA condition achieved an average accuracy of 62.7\%, compared to 41.3\% in the Manual condition—a statistically significant increase ($p<0.001$). When AI assistance was combined with knowledge provision (KA+AIA), the average accuracy further increased to 64.4\%.

Interestingly, while knowledge provision alone (KA) improved average accuracy to 43.8\% compared to the Manual condition, this difference was not statistically significant ($p=0.127$). This suggests that AI assistance has a more substantial effect on improving defect detection accuracy than knowledge provision alone.

Furthermore, the interquartile range (IQR) for participants in the AI-assisted conditions was narrower than in the Manual and KA conditions, indicating more consistent performance among participants when AI assistance was provided. This consistency likely results from the AI system's ability to reduce cognitive load by automating the detection of surface-level defects, allowing human testers to focus on more complex issues.

These findings demonstrate that AI assistance significantly enhances human testers' effectiveness in identifying game defects, and that combining AI assistance with detailed knowledge yields the highest accuracy. The results indicate the potential of AI to augment human capabilities in game testing, leading to more efficient and accurate defect detection processes.

\subsection{Impact of Gaming Experience on Testing Accuracy}

Participants without prior gaming experience showed a similar boost in accuracy from AI assistance, achieving 61.4\% with AI and 45.7\% without AI when no extra knowledge was provided (AIA vs. Manual). However, with the addition of knowledge, accuracy decreased to 59.3\% with AI and 35.0\% without AI (KA+AIA vs. KA). This suggests that while AI consistently improves accuracy, unfamiliarity with video games may make knowledge provision overwhelming, reducing effectiveness. Despite this, AI assistance helped this group achieve accuracy comparable to participants with prior gaming experience.

Given the gaming scenario, we attracted few non-gamers. Due to the small sample size of 7 participants in this group, these findings are not statistically significant. Future work should involve a more diverse participant pool to better understand how gaming experience influences the effectiveness of AI and knowledge in game testing.

\subsection{Accuracy of AI-Assisted Systems Under Challenging Scenarios}\label{EASUCC}

To further assess the value of AI assistance, we analyzed participants' accuracy in identifying defects under challenging conditions, such as low visibility (LV), multiple defects (MD), subtle defects (SD), and ambiguous boundaries (AB). Table~\ref{tab3} summarizes the accuracy across different conditions.

\begin{table}
\caption{Accuracy of AI-Assisted Systems Under Challenging Conditions}
    \centering
    \begin{tabular}{|c|c|c|c|c|} \hline 
         & Manual (\%) & AIA (\%) & KA (\%) & KA+AIA (\%) \\ \hline 
         LV &  42.4 &  71.6 &  47.0 &  65.3\\ \hline 
         MD &  44.0 & 67.9 & 48.3 &  65.6\\ \hline 
         SD & 36.6 & 54.7 & 46.5 & 62.3\\ \hline
         AB & 36.2 & 65.4 & 46.6 & 72.9\\ \hline
         Others & 41.3 & 62.9 & 43.0 & 64.5\\ \hline
    \end{tabular}
    \label{tab3}
\end{table}

\paragraph{Low Visibility (LV)} In low-visibility conditions, manual testers achieved an accuracy of 42.4\%, indicating the difficulty of detecting defects when visibility is compromised. AI assistance substantially improved accuracy to 71.6\% in the AIA condition ($p<0.001$ compared to Manual), demonstrating AI's effectiveness in processing visual information that may be challenging for humans. However, the KA+AIA condition saw a slightly lower accuracy of 65.3\%, suggesting that additional knowledge may not significantly enhance performance in conjunction with AI assistance under low-visibility conditions, possibly due to information overload or cognitive interference.

\paragraph{Multiple Defects (MD)} When test cases contained multiple defects, manual testers achieved 44.0\% accuracy. AI assistance again led to a significant improvement, with 67.9\% accuracy in the AIA condition ($p<0.001$). The KA+AIA condition achieved 65.6\% accuracy, comparable to the AIA condition. This indicates that AI assistance helps testers manage the complexity of identifying multiple defects simultaneously, likely by systematically highlighting potential issues.

\paragraph{Subtle Defects (SD)} Detecting subtle defects proved challenging for manual testers, with an accuracy of 36.6\%. AI assistance improved accuracy to 54.7\% in the AIA condition ($p<0.001$), and further to 62.3\% in the KA+AIA condition. The significant improvement in the condition KA + AIA suggests that the combination of AI assistance and detailed knowledge enables testers to detect subtle defects more effectively, perhaps by providing context that enhances the interpretation of AI output.

\paragraph{Ambiguous Boundaries (AB)} In cases with ambiguous defect boundaries, manual testers had the lowest accuracy (36.2\%). The AIA condition improved accuracy to 65.4\% ($p<0.001$), while the KA+AIA condition achieved the highest accuracy of 72.9\%. This substantial improvement indicates that when defects are difficult to classify, the combination of AI assistance and knowledge provision is particularly beneficial, likely because the detailed information helps testers resolve ambiguities in defect categorization.

Overall, these results highlight that AI assistance significantly enhances defect detection accuracy under challenging conditions, and that the addition of knowledge provision can further improve performance in certain scenarios. The findings emphasize the importance of integrating AI tools with human expertise to address complex testing challenges effectively.


\subsection{Impact of AI Descriptive and Judgment Errors on Human Participants}

While AI assistance generally improved defect detection accuracy, it is crucial to understand how AI errors affect human testers. We analyzed the impact of AI descriptive errors (DE), judgment errors (JE), cases with correct description but judgment errors (CDJE), and cases with both description and judgment errors (DEJE) on participant accuracy. Table~\ref{tab4} presents the accuracy across different conditions.

\begin{table}
\caption{Impact of AI Error Information on Human Participants}
    \centering
    \begin{tabular}{|c|c|c|c|c|} \hline 
         & Manual (\%) & AIA (\%) & KA (\%) & KA+AIA (\%) \\ \hline 
         DE &  41.5 &  35.2 &  38.2 & 44.5 \\ \hline 
         JE & 32.2 & 20.3 & 27.0 & 27.4 \\ \hline 
         CDJE & 37.0 & 51.7 & 25.5 & 54.6 \\ \hline 
         DEJE & 30.6 & 9.8 & 27.5 &  18.4 \\ \hline
         Others & 41.3 & 68.0 & 45.5 & 68.5 \\ \hline
    \end{tabular}
    \label{tab4}
\end{table}

\paragraph{Descriptive Errors (DE):} In cases where the AI provided incorrect scene descriptions, participant accuracy decreased in the AIA condition (35.2\%) compared to the Manual condition (41.5\%). This suggests that misleading descriptions from the AI can negatively impact testers who rely on AI outputs. However, in the KA+AIA condition, accuracy improved to 44.5\%, indicating that access to detailed knowledge helps participants identify and correct AI descriptive errors.

\paragraph{Judgment Errors (JE):} When the AI misclassified defects, participant accuracy was significantly lower in the AIA condition (20.3\%) compared to the Manual condition (32.2\%), highlighting the potential for AI judgment errors to mislead testers. The KA+AIA condition showed a slight improvement (27.4\%), suggesting that knowledge provision partially mitigates the impact of AI judgment errors by enabling testers to critically assess AI outputs.

\paragraph{Correct Description, Judgment Error (CDJE):} In scenarios where the AI provided accurate descriptions but incorrect defect judgments, participants in the AIA condition achieved higher accuracy (51.7\%) than in the Manual condition (37.0\%). This implies that accurate descriptions allow testers to override incorrect AI judgments. The highest accuracy was observed in the KA+AIA condition (54.6\%), reinforcing the benefit of combining AI assistance with knowledge provision.

\paragraph{Both Description and Judgment Errors (DEJE):} When the AI made both descriptive and judgment errors, participant accuracy was notably low in the AIA condition (9.8\%), significantly lower than in the Manual condition (30.6\%). This shows the detrimental effect of compounded AI errors on human performance. The KA+AIA condition showed some improvement (18.4\%), indicating that knowledge provision can help testers identify and correct AI errors to some extent.

These findings highlight that while AI assistance generally enhances performance, AI errors can adversely affect human testers, particularly when they rely heavily on AI outputs. Providing testers with additional knowledge and training to critically evaluate AI outputs is essential to mitigate the impact of AI errors.

\subsection{Impact of AI Hallucinations on Bug Detection}



AI hallucinations—instances where the AI incorrectly identifies defects in non-defective visuals or fails to detect actual defects—pose a significant challenge in AI-assisted game testing. We examined how AI hallucinations affected participant accuracy, focusing on false negative detections (FND) and false positive detections (FPD). Table~\ref{tab5} summarizes the results.

\begin{table}
\caption{Accuracy When AI Hallucinates Bugs}
    \centering
    \begin{tabular}{|c|c|c|c|c|} \hline 
         & Manual (\%) & AIA (\%) & KA (\%) & KA+AIA (\%) \\ \hline 
         FND &  34.8 &  24.6 &  27.4 & 35.7 \\ \hline 
         FPD & 27.3 & 26.3 & 21.5 & 21.8 \\ \hline 
    \end{tabular}
    \label{tab5}
\end{table}

\paragraph{False Negative Detections (FND):} In cases where the AI failed to detect actual defects, participants in the AIA condition had lower accuracy (24.6\%) compared to the Manual condition (34.8\%). This indicates that reliance on AI can lead to missed defects when the AI overlooks issues. However, the KA+AIA condition showed an accuracy of 35.7\%, similar to the Manual condition, suggesting that knowledge provision helps participants compensate for AI's false negatives by relying on their own judgment informed by detailed information.

\paragraph{False Positive Detections (FPD):} When the AI incorrectly identified defects in non-defective visuals, participant accuracy was low across all conditions. The Manual condition achieved slightly higher accuracy (27.3\%) compared to the AIA (26.3\%) and KA+AIA (21.8\%) conditions. This suggests that AI hallucinations of false positives can mislead testers, even when additional knowledge is provided. Participants may trust the AI's assessment, especially if the AI presents plausible explanations for the supposed defects, leading to incorrect defect reports.

\section{Discussion}


In our analysis, one of the key outcomes is that participants utilizing AI assistance consistently achieved higher accuracy compared to those relying solely on manual testing. As shown in Table~\ref{tab3}, AI-assisted conditions (AIA and KA+AIA) led to substantial improvements in defect detection across scenarios with low visibility, multiple defects, subtle defects, and ambiguous boundaries. This demonstrates the potential of AI systems to augment human capabilities by efficiently handling surface-level defect detection, allowing testers to focus on more complex issues.

However, the study also indicates that AI assistance is not infallible. AI errors—both descriptive and judgmental—negatively impacted human performance. When the AI provided incorrect descriptions or misclassified defects, participants' accuracy decreased significantly in the AI-assisted conditions (Table~\ref{tab4}). Moreover, AI hallucinations, where the AI incorrectly identified non-defective visuals as containing bugs or failed to detect actual defects, further misled testers (Table~\ref{tab5}). These findings align with existing literature on the risks of over-reliance on AI systems, where users may accept AI outputs without sufficient critical evaluation \cite{rawte2023survey,zhang2023siren}.

Importantly, combining AI assistance with detailed knowledge provision (KA+AIA) mitigated some of the negative effects of AI errors. Participants with access to additional knowledge were better equipped to critically assess AI outputs, correct errors, and maintain higher accuracy levels, even when the AI made mistakes. For instance, in cases of correct description but judgment errors (CDJE), the KA+AIA condition achieved the highest accuracy (54.6\%), indicating that knowledge provision helps testers override incorrect AI judgments (Table~\ref{tab4}). This emphasizes the crucial role of human oversight and the value of equipping testers with sufficient context to make informed decisions.

The study also reveals that while knowledge provision alone (KA) did not significantly improve accuracy compared to manual testing, its combination with AI assistance led to the highest performance. This suggests that AI assistance enhances human performance most effectively when testers are supported by both AI tools and adequate knowledge resources. However, there is a need to balance the amount of information provided to avoid cognitive overload, which can hinder decision-making efficiency \cite{andrejevic2013infoglut,zou2024docbench}.

Our findings highlight the importance of designing AI systems that are not only accurate but also transparent and interpretable, enabling testers to understand and critically evaluate AI outputs. Additionally, training testers to maintain a critical perspective towards AI assistance and to effectively use the provided knowledge resources is essential to maximize the benefits of AI integration.

\section{Conclusion}

Through a comprehensive user study, we demonstrated that AI assistance significantly enhances real-world human testers' effectiveness in identifying game defects, particularly when combined with detailed knowledge resources. By integrating AI-driven defect detection with human oversight, we experimented a hybrid testing workflow that optimizes efficiency and accuracy in game testing processes.

However, the research also highlights challenges associated with AI errors and hallucinations, which can negatively impact performance if not properly managed. These findings emphasize the critical need for effective human-AI collaboration, where testers are equipped to critically assess and correct AI outputs. Developing improved collaborative frameworks and interaction models is essential to address the problems associated with AI errors and to facilitate better and easier human judgment.

Our study contributes to the understanding of human-AI interaction in complex decision-making environments, revealing the dual role of AI as both an enhancer and a potential source of error in game testing. Practically, the study suggests that providing testers with manageable knowledge resources and training in critical evaluation can maximize the benefits of AI assistance while mitigating risks.

Limitations of this study include a participant pool with predominantly gaming experience, which may affect generalizability, and a focus on specific defect types and testing conditions. Future research should include a more diverse range of participants and expand the scope to various game genres and complex scenarios. Additionally, improving AI transparency and developing strategies to enhance human oversight will be crucial for advancing AI-assisted game testing methodologies.

In conclusion, by addressing the identified challenges and fostering effective human-AI collaboration, the gaming industry can leverage AI to achieve more efficient, accurate, and reliable testing processes.

\bibliographystyle{IEEEtran}
\bibliography{IEEEabrv,main}

\end{document}